# Residue Network Construction and Predictions of Elastic Network Models

**Canan Atilgan, Ibrahim Inanc, and Ali Rana Atilgan**

Faculty of Engineering and Natural Sciences, Sabanci University, 34956 Istanbul, Turkey

*e-mail*: canan@sabanciuniv.edu
*telephone*: +90 (216) 4839523
*telefax*: +90 (216) 4839550

**ABSTRACT**

The past decade has witnessed the development and success of coarse-grained network models of proteins for predicting many equilibrium properties related to collective modes of motion. Curiously, the results are usually robust towards the different methodologies used for constructing the residue networks from knowledge of the experimental coordinates. We present a systematical study of network construction strategies, and their effect on the predicted properties. The analysis is based on the radial distribution function and the spectral dimensions of a large set of proteins as well as a newly defined quantity, the angular distribution function. By partitioning the interactions into an essential and a residual set, we show that the robustness originates from a large number of long-distance interactions belonging to the latter. These residuals have a vanishingly small effect on the force vectors on each residue. The overall force balance then translates into the Hessian as small shifts in the slow modes of motion and an invariance of the corresponding eigenvectors. Implications for the study of biologically relevant properties of proteins are discussed.

**INTRODUCTION**

Globular proteins show diversified structures and sizes, yet, it has been claimed that they display a nearly random packing of amino acids with strong local symmetry on the one hand (1), and that they are regular structures that occupy specific lattice sites, on the other (2). It was later shown that this classification depends on the property one investigates, and that proteins display "small-world" properties, where highly ordered structures are altered with few additional links (3). Furthermore, packing density of proteins scales uniformly with their size (4, 5) which causes them to show similar vibrational spectral characteristics to those of solids (6).

Dynamical studies of folded proteins draw much attention to their importance in relating the structure of the proteins to their specific function and collective behavior. Protein dynamics is generally both anisotropic and collective. Internal motional anisotropy is a consequence of the general lack of symmetry in the local atomic environment, while the collectivity is mainly caused by the dense packing of proteins (7).

Theoretical studies on fluctuations and collective motions of proteins are based on either molecular dynamics (MD) simulations or normal mode analysis (NMA). Since, in molecular simulations with conventional atomic models and potentials, computational effort is demanding for larger proteins with more than a few hundreds of residues, coarse grained protein models with simplified governing potentials have been employed, and these have shown a great success in the description of the residue fluctuations and the collective behavior of proteins (8).

One of these simplified models, NMA using a single parameter harmonic potential (9) successfully predicts the large amplitude motions of proteins in the native state (10). Within the framework of this model, proteins are modeled as elastic networks whose nodes are residues linked by inter-residue potentials that stabilize the folded conformation. The residues are assumed to undergo Gaussian-distributed fluctuations about their native positions. The springs connecting each node to all other neighboring nodes are of equal strength, and only the atom pairs within a cut-off distance are considered without making a distinction between different types of residues. This model, with its simplicity, speed of calculation and relying mostly on geometry and mass distribution of the protein, demonstrates that a single-parameter model can reproduce complex vibrational properties of macromolecular systems. By separating different components of normal modes, e.g collective (low-frequency) motions, the nature of a conformational change, for example due to the binding of a ligand, can also be analyzed thoroughly (11).

Following the uniform harmonic potential introduced originally by Tirion (12), residue level application of elastic network models paved the way for the concept Gaussian Network Model (GNM), which is based on the energy balance of the system at the energy minimum, and is a purely thermodynamic treatment (10, 13). Elastic models based on the force balance around each node (14) led to the development of the so called Anisotropic Network Model (ANM) (15). In the past few years, variant methods of GNM and ANM (16, 17) have been introduced. The applications of these models to many proteins show successful results in terms of predicting the collective behavior of proteins. Despite numerous applications comparing the theoretical and experimental findings on a case-by-case basis (18-22), only a few attempted a statistical assessment of the models. A methodology that evaluates the number of modes necessary to map a given conformational change from the degree of accuracy obtained by the inclusion of a given number of modes, showed the results to be protein dependent (23). In another study where 170 pairs of structures were systematically analysed, it was shown that the success of coarse-grained elastic network models may be improved by recognizing the rigidity of some residue clusters (24).

To date, the structures that form the basis of the network models have been generated from certain rules of thumb. In GNM, which does not include directionality and is therefore a one-dimensional model, the first correlation shell between the $C_\alpha$ or $C_\beta$ atoms of the residues is used as the rule for the connectedness of a given pair of residues (ca. $6.7 - 7.0$ Å) (10). In the three-dimensional ANM, values in the range of $8 - 14$ Å are found in the literature based on the argument that (i) the eigenvalue distributions obtained from the modal decomposition are similar to those obtained from the full-atom NMA description of proteins, or (ii) these provide atomic fluctuation profiles that display the largest correlation with the experimental B-factors. Voronoi tessalation of the space defined by the central (usually $C_\alpha$ or $C_\beta$) atom into non-intersecting polyhedra constitute another route that frees one from defining a cut-off distance (25). Atom-based network construction approaches have also been used [see reference (26) for a review of the variety of network construction methods in literature.]

In this study, we use a systematic approach on a large set of globular proteins with varying architectures and sizes to find a basis for why the network models work well to define certain properties of the system. This enables us to assess the various residue-based approaches used in the construction of the networks. We define a direction based radial distribution function for this purpose, and show that the directionality of newly added links samples a spherically symmetric collection of directions beyond a given distance of interacting residues. We show that the network construction is free of the cut-off distance problem once a certain baseline threshold is accessed, if one is interested in the collective motions and the fluctuation patterns of the residues. Implications for the limitations of the ANM methodology are discussed due to functionality-related predictions based on the most global motions.

## COMPUTATIONAL DETAILS

**Network construction.** A protein of $N$ residues is treated as a residue-based structure, where the $C_\alpha$ atom of each amino acid is considered as a node, and the coordinates of the protein are obtained from the protein data bank (PDB) (27). The network information is contained in the $N \times N$ adjacency matrix, **A**, of inter-residue contacts, whose elements $A_{ij}$ are taken to be 1 for contacting pairs of nodes $i$ and $j$, and zero otherwise. We determine the presence of a contact using two approaches, one involving a selected cut-off distance, and the other using Voronoi tessellations. In the former approach, the criterion for contact is that the two nodes are within a cut-off distance $r_c$ of each other. In the latter, Voronoi cells are formed from the PDB coordinates of $C_\alpha$ atoms such that the three dimensional space is uniquely and completely subdivided into polyhedra whose surfaces are defined by the intersection of contact planes built midway between the nodes of the network. Thus, pairs of nodes sharing a common plane are taken to be in contact. This methodology allows eliminating the choice of a cut-off distance so that an unambiguous network construction is achieved (1, 28). We have utilized the freely available Voro3D program for this purpose (25). Note that the nodes in these networks have an average distance of 6.6 Å to their neighbors, and an average contact number of 10.5.

**Radial and angular distribution functions.** The radial distribution function (RDF), $g(r)$, is a measure of the correlation between the locations of particles within a system, measured as the probability of finding another particle at a distance, $r$, from a chosen particle, normalized by the volume element. The RDF is a useful tool to describe the structure of a molecular system, particularly those of liquids. In an ordered solid, RDF has an infinite number of sharp peaks whose separations and heights are characteristic of the lattice structure. RDF can be deduced experimentally from X-ray or neutron diffraction studies, thus providing a direct comparison between experiment and simulation.

In the current work, we are not only interested in the number distribution of particles around a given node, but also concentrate on the link structure. We treat all neighbors of a node equivalently, and we find that as $r_c$ is increased with the addition of new neighbors to each node, the resultant vector, $\mathbf{Q}_i$, on node $i$ due to all its neighbors, $j$, converges to a certain location:

$$Q_i = \sum_j A_{ij} R_{ij} \qquad (1)$$

where $\mathbf{R}_{ij}$ is the unit vector connecting residue pairs $i$ and $j$, and $A_{ij}$ are the elements of the adjacency matrix. An example is shown on a 54 residue α-helical protein (PDB code: 1enh) in figure 1, where the length of a red vector is proportional to $r_c$ and demonstrate that at small $r_c$, the neighbors of a node are at distinct locations, whereas with increasing $r_c$, the new nodes are added in a spherically symmetrical manner so that the resultant vector, $\mathbf{Q}_i$, is only slightly modified. The resultant vectors from the Voronoi tessalated network structure is also shown (in yellow) and is found to be different from the converged ones.

To quantify this behavior, we define the angular distribution function (ADF), which is the distribution of angular change, $\Delta\varphi$, of the resultant vector obtained from the contacting residues at a distance $r$ to $r+dr$ to the reference residue:

$$\cos\Delta\varphi_i(r) = \left(\sum_j A_{ij}R_{ij}\right)_r \cdot \left(\sum_j A_{ij}R_{ij}\right)_{r+dr} = Q_i|_r \cdot Q_i|_{r+dr} \qquad (2)$$

where $dr$ is a small perturbation on the distance $r$.

**Anisotropic network model.** In the anisotropic network model (ANM)**,** the networks are formed as described under the subsection *Network Construction* and the interactions between nodes is considered to be due to harmonic potentials (15). Nodes within the predetermined cut-off distance $r_c$ are coupled by elastic springs having a uniform force constant $\gamma$. Thus, the overall potential of the molecule is given by the sum of all harmonic potentials among interacting nodes such that

$$V = \frac{\gamma}{2} \sum_i \sum_{j>i} (A_{ij})(R_{ij} - R_{ij}^0)^2 \,. \qquad (3)$$

$R_{ij}^0$ is the average distance between residues $i$ and $j$. For a network of $N$ nodes, the Hessian is a $3N \times 3N$ matrix formed by a number of $N^2$ super elements $\mathbf{H}_{ij}$. The off-diagonal super elements of $\mathbf{H}_{ij}$ ($i \neq j$), obtained from the second derivative of the total potential with respect to node positions, are given by

$$\mathbf{H}_{ij} = \frac{\gamma A_{ij}}{(R_{ij}^0)^2} \begin{bmatrix} X_{ij}X_{ij} & X_{ij}Y_{ij} & X_{ij}Z_{ij} \\ Y_{ij}X_{ij} & Y_{ij}Y_{ij} & Y_{ij}Z_{ij} \\ Z_{ij}X_{ij} & Z_{ij}Y_{ij} & Z_{ij}Z_{ij} \end{bmatrix}, \qquad (4)$$

where $X_{ij}$, $Y_{ij}$, and $Z_{ij}$ are the Cartesian components of the distance vector $R_{ij}^0$. The diagonal super elements are given by $\mathbf{H}_{ii} = -\sum_{j, j \neq i} \mathbf{H}_{ij}$. An equivalent way to forming the Hessian is based on the force balance around each node (14, 15), and involves the direction cosines along these vectors; this issue will be taken up in detail in the Appendix.

The pseudo-inverse of $\mathbf{H}$ is the $3N \times 3N$ covariance matrix, $\mathbf{C}$, that can be expressed in terms of the $3N-6$ non-zero eigenvalues $\lambda_k$ and corresponding eigenvectors $\mathbf{u}_k$ of $\mathbf{H}$ as

$$\mathbf{C} = \sum_{k=1}^{3N-6} \frac{1}{\lambda_k} \mathbf{u}_k \mathbf{u}_k^T.$$

(5)

Here, the eigenvectors $\mathbf{u}_k$ represent the spatial dependence (direction) of each mode $\lambda_k$. The smallest nonzero eigenvalue $\lambda_1$ corresponding to the lowest frequency is assumed to carry information on the most collective internal modes of motion.

The residue fluctuations are predicted by the ANM for residue $i$ from the trace of $\mathbf{C}_{ii}$. Theoretically, they are related to the B-factors determined from X-Ray crystallografic data through the relation,

$$\mathbf{B}_i = (8\pi^2 k_B T / 3\gamma) \, tr(\mathbf{C}_{ii})$$

(6)

where $k_B$ is the Boltzmann constant and $T$ is the absolute temperature. The value of $\gamma$ is determined *a posteriori* if experimental data are available, and does not affect the fluctuation profile of residues.

**Protein data sets.** We base our calculations on a set of 595 proteins with sequence homology less than 25% and sizes spanning 54–1021 residues (29). This protein set is identical to that used in our previous statistical analyses on residue networks (3, 30). Forty-five of the proteins in the set have fewer than 100 residues, the number of proteins in the ranges (101–200), (201–300), (301–400), and more than 400 residues are 234, 122, 108, and 86, respectively. A list of all the proteins used, their sizes, and distributions appear in the Supplementary Material of reference (30).

In addition, we have studied the location dependence of certain properties. For this reason, we calculate residue depth from the surface of the protein (31, 32). We classify residues that are deeper than 4 Å as core, and the rest of them as surface residues. The choice of this value is based on the fact that the size of spatial fluctuations, as calculated from MD simulations on BPTI, of the surface and interior residues converge to the same value at the protein dynamical transition (33). For the distinction of core/surface residues, we use a subset of the original protein data set that has a total of 60 representatives with sizes in the range 140 − 320 amino acids. Finally, we also study the eigenvalue spectra of proteins ($\lambda_k$ in equation 5), which is affected by the size of the systems. We therefore choose the subset of 26 proteins for which $N$ = 150 ± 10.

## RESULTS AND DISCUSSION

**Structural heterogeneity of amino acid distributions in proteins.** The RDF, $g(r)$, of the residues is presented in Figure 2a for distances up 20 Å, recorded at 0.1 Å resolution. We find that the first sharp peak in $g(r)$ ends at ca. 6.7 Å corresponding to the first coordination shell (i.e., the range within which residue pairs are found with the highest probability), the second coordination shell occurs at 8.5 Å. Broader peaks ending at 10.5 and 12 Å are identified as the third and fourth coordination shells. At larger distances, $g(r)$ monotonically decreases, indicating that the coarse-grained residue beads do not experience further ordering in the liquid-like environment. In Figure 2a we also display the ADF, $g(\varphi)$, for the same set of proteins in the same distance range. We find that the main peaks of ADF and RDF overlap, the only difference in the general character of the two distribution functions being found in the third and fourth coordination shells. In RDF, we find that a similar number of particles per unit volume exist in these two coordination shells (same height in the distribution). The ADF provides the additional information that, due to the asymmetry in the intensities of the third and fourth coordination shells, these particles are clustered in relatively more ordered directions in the third shell, quantified by the increase in ADF to ca. 5°. The ADF provides the valuable information that the additional particles are taken into account as more concentric spherical shells of 0.1 Å diameter are added (recall figure 1), have a preferred direction of clustering at the regions of higher number density. Conversely, at larger distances, the new neighbors carry directionality that cancel each other out, as would be expected from a random packing of spheres, quantified by the monotonical decrease in $g(\varphi)$.

Since globular proteins may be considered to be made up of a core region surrounded by a molten layer of surface residues (34), it is of interest to distinguish the topological differences between the core and the surface (Figure 2b). We observe that core residues have larger angular changes in the resultant vector, $\mathbf{Q}_i$ (equation 1) compared to the surface residues. Note that the fraction of surface residues is ca. 0.6 for these proteins, being somewhat larger for the smaller sized ones (3). Thus, the resultant vector on the surface residues rapidly converges to a given directionality specific to each residue at short distances, the additional links at higher distances arriving in directions that cancel out. The overall structural heterogeneity is detected much clearly in the $g(\varphi)$ of the core residues. However, the heterogeneity in the first coordination shell is more pronounced over that of the second for the surface residues, possibly due to the loose packing in this region. This effect is reversed in the core. In addition, the structural asymmetry between the third and fourth coordination shells is found to originate from the structure of the core residues. The dissimilar behavior of the core and surface regions is also observed in Figure 1b. As $r$ increases, the orientation of the vectors are more scattered in the interior, indicating its isotropic nature; conversely, the orientation of the vectors at the surface rapidly converges.

**Density of vibrational normal modes.** The vibrational normal mode spectra, $g(\omega)$, of proteins was originally studied by ben-Avraham for five proteins with sizes in the range of 39 − 375 residues, the data collapsing on a single curve, especially in the slow region (6). The density of states was found to increase linearly with the frequency in this region, implying a spectral dimension of $d_s = 2$ and deviating from the Debye model of elastic solids where the expected value is 3 (35). The anomalous spectral dimensions of proteins was also confirmed by inelastic neutron scattering experimental measurements, which yielded $d_s \approx 1.4$ for hen egg white lysozyme (36). More recently, an equation of state relating the spectral dimension, fractal dimension and the size of a protein was developed based on the coexistence of stability and flexibility in folded proteins (37).

In the original ANM study, the cut-off distance used in network construction was roughly chosen to mimic this distribution of the modes (15), which was 13 Å for the retinol binding protein studied therein; however, a wide range of cut-off distances appear in the literature based on other criteria, as discussed in the Introduction. Nevertheless, constructing networks with harmonic potentials whose spectra closely mimic the vibrational modes from all-atom systems seems to be the most plausible approach, since this implies that the curvatures of the energy functions used in the two approaches are adequately approximated, so that the equilibrium properties would be described properly.

In figure 3a, we display the $r_c$ dependence of normal mode spectra averaged over 26 proteins of size $150 \pm 10$ residues, enabling us to disregard the size effect in the calculations [the latter was addressed in references (37) and (38).] In general, the low-frequency band of the graph is responsible for large amplitude collective motions related to function, whereas the high-frequency band refers to small amplitude motions of individual residues. We find that at $r_c = 7$ Å (where neighbors are from the first coordination shell), the distribution is characterized by a direct drop in density with increasing frequency; at this value, most proteins have additional zero eigenvalues, apart from the six due to the rigid body motions. The universal behavior of the slow vibrational modes of proteins is recovered at higher $r_c$ values. Above the cut-off distances that include the fourth coordination shell ($r_c > 12$ Å), a shoulder in the higher frequency region first appears, then broadens as $r_c$ is increased. At $r_c > 16$ Å, a two-peaked density profile that is uncharacteristic of proteins sets in (inset to figure 3a). For the networks obtained with Voronoi tessalations (dashed line in figure 3a), the distribution shows a flat behavior, also uncharacteristic of proteins. Also note that, although the average distance between adjacent nodes is 6.6 Å in these systems, their behavior is markedly different from that of the networks with similar cut-offs (e.g. $r_c = 7$ Å.)

Thus, an $r_c$ value in the range of $8 - 16$ Å captures the general shape of protein vibrational spectra. Yet, inasmuch as one utilizes network models to study collective motions of proteins as a superposition of several low frequency modes, it is important to capture the distribution in the slow mode region of the protein in more detail. This region is intimately related to material properties, characterized by the spectral dimension, $d_s$. In figure 3b, we plot the spectral dimensions of these systems, obtained from power law best-fits to the cumulative density of modes, $G(\omega) \propto \omega^{d_s}$ for the first 70 modes in each set of data [with $dG(\omega)/d\omega = g(\omega)$]. The dimensions approach the Debye model value of 3 as $r_c$ is increased (dotted line in the figure). The spectral dimension of the Voronoi tessalated networks is 1.0, and is commensurate with that of the network at $r_c = 9$ Å. The spectral dimensions in the $r_c$ range from the second to the fourth coordination shell, ($8 - 12$ Å increase from below $d_s = 1$ to ca. $d_s = 1.5$. Furthermore, a crossover in the rate of change of the spectral dimension with the cut-off distance occurs at $r_c = 16$ Å, the slope reducing from ca. 0.13 to half this value; the crossover is accompanied by the shift to $d_s > 2$. Thus, it is plausible to use the cut-off value up to 16 Å so as to capture both the general shape of the vibrational spectra of proteins, as well as the spectral dimension that describes the density of slow modes.

**Biological significance.** In recent years, network models of proteins, RNA and their complexes have opened up previously unprecedented areas of study, since the level of coarse graining adopted has been shown to describe several important phenomena unique to these self-assembled systems. The findings are mainly based on the observation that a simplified harmonic potential (equation 3) is capable of describing the collective modes of motion (7), which also are associated with the basic functioning of these molecular machines (13). First, it was demonstrated that the Debye-Waller factors obtained from X-ray crystallography correlate with the fluctuations predicted by the theory (10). This led to the study of the cross-correlations between the different parts of the system with confidence, leading to information not directly accessible by experiments; in particular, the coupled motions in the low frequency regions were found to shed light on many experimental findings and were utilized to uncover some mechanistic features; see, e.g. (39). It was later shown that the eigenvectors associated with the lowest frequencies of motion also described the conformational changes accompanying binding (40, 41). The level of success in the latter work depends on the degree of collectivity displayed by the particular protein (24), and the number of modes that describe the essential motions is highly specific to the protein, or even to the different ligand bound forms of the same protein (23). Note that such analyses require knowledge of both bound and unbound forms of the protein for proper assessment of the predictions. Nevertheless, the eigenvectors may effectively be used to e.g. generate unique starting structures for MD simulations to perform higher level calculations. Coupled with structural alignment tools, they may also hint at relationships between enzymes that otherwise lack local or global structural similarities (42).

The level of success of these studies in relation to the method of network construction has not been addressed systematically. We find for a number of proteins that the correlation between the mean-square fluctuations of $C_\alpha$ atoms and the theoretical predictions of equation 6 improve as the cut-off distance is increased. This curious observation is valid up to very large $r_c$ values; i.e. for some proteins, even when all residues are interconnected, the fluctuations of individual residues are faithfully predicted. One example is displayed in figure 4 for a 263 residue $\beta$-class protein (PDB code: 1arb), where the residue-by-residue experimental B-factors (middle curve in gray in Figure 4a) are compared with several selected theoretical models: A relatively low correlation is obtained at $r_c = 8$ Å; in particular, the fluctuations of surface loop residues $15 - 20$ and $135 - 145$ are overestimated due to the absence of important core-region contacts that are not taken into account at this cut-off distance. The $r_c = 15$ Å model captures the experimentally determined fluctuation patterns, which remains unaltered at higher cut-offs. The fluctuations predicted by the Voronoi tessalated network model are somewhat chaotic, lowering the correlations with experiment. The Pearson correlation coefficients at a wide range of cut-off distances are plotted in figure 4b, along with the value obtained from Voronoi tessalated networks (dashed line). We emphasize that the behavior exemplified by figure 4 is not unique to this protein, but is rather a common property of all proteins.

The increase in the correlation coefficient with $r_c$ as well as its persistence to very high $r_c$ values implies that the main ingredients that contribute to the fluctuation predictions are present in the Hessian obtained at a relatively low $r_c$, and the additional contacts act as a perturbation to this "essential" part of the matrix. We may thus partition the Hessian into two (equation A1), where $H_*$ contains information due to the essential contacts of the matrix whereas $H_r$ is the residual part where the interactions are added in a spherically symmetrical manner around the nodes beyond a certain $r_c$ value (figure 2). In the Appendix we include a proof of how the lowest eigenvalues are modified in a small window based on this partitioning, as well as how the eigenvectors remain unchanged (equation A7). The result in equation A8 implies that the inverse of the Hessian will be nominally modified, so that the predicted $C_\alpha$ fluctuations will change only slightly. On the other hand, since a Voronoi tessalated network construction of proteins is based on the minimal partitioning of space in the neighborhood of nodes, it will not form the essential part of the Hessian, $H_*$, which must necessarily include a larger number of contacts from the first coordination shell.

Finally the findings outlined in the Appendix imply that, due to the invariance of the eigenvectors under such a perturbation to the essential part of the Hessian, the mode based predictions on the direction of motion between the unbound and bound conformations of the protein will also converge. An example is shown in figure 5 for the protein adenylate kinase, for which the eigenvector that belongs to the lowest eigenvalue is known to describe the conformational change with high accuracy due to the highly collective behavior of the hinge motion between the two domains (43). The Pearson correlation between the experimental and theoretical curves is 0.9 at all cut-off distances above $r_c = 8$ Å. The largest discrepancy between theory and experiment is observed in the region spanning residues $30 - 67$ which belongs to the NMP binding domain closing over the ATP binding domain (called the LID) on the opposite side, the latter spanning residues $118 - 167$. The prediction does not change with $r_c$ and may possibly be improved by the inclusion of additional modes, which is beyond the scope of the current work.

To recapitulate, with our analysis over a large set of non-homologous proteins, the degree of success of network models of proteins is shown to converge as the cut-off distance used in constructing the network from the PDB coordinates of the protein is increased. A choice of high $r_c$ in the vicinity of 16 Å covers the neighborhood structure of an arbitrary protein and its eigenvalue spectra; however, for large proteins, this will introduce a large number of interactions which will render the matrix inversion procedure rather cumbersome. In such cases, one may resort to compute $g(r)$, $g(\varphi)$ and $g(\omega)$ curves and spectral dimensions for the particular protein to choose an optimum $r_c$; for large proteins the number of nodes will be high enough to obtain statistics for smooth curves where the peaks may be discerned, a problem that cannot be circumvented for small system sizes. We note that network models are useful in describing the properties related to the fluctuations near the minimum of the conformational energy well, and its curvature. However, they will not succeed in providing information of the dynamical properties of the protein, unless a methodology for updating the Hessian along the reaction coordinate is introduced.

## CONCLUSIONS

Despite their different topological structures and sizes, a statistical analysis of a large number of folded proteins leads to common features. In particular, the radial and angular distribution functions provide the degree of (in)homogeneity in the protein as well as a quantitative description of the location of the coordination shells. Depth dependent analysis shows that the densely packed core region of the protein has a different local structure built around it compared to its surface. In the core of the protein, the second neighbors have a non-random distribution that is more pronounced than the first neighbors. In the surface residues, the reverse is observed (figure 2).

Calculations at a variety of cut-off distances used in network construction reflect that the dimensionality of the system approaches that of regular crystals where $g(\omega)$ scales with $\omega^2$ only at unrealistically high $r_c$ values (figure 3b). The modal spectrum resembles that obtained from all-atom calculations with realistic atom-atom interaction potentials in the region above the second coordination shells up to a cut-off distance of 16 Å (figure 3a). At this threshold, the spectral dimension shifts from the region of $d_s = 1$-$2$ to above 2, accompanied by a crossover in its rate of change (figure 3b).

We have shown that the slow modes are immune to the details of network construction once the essential contacts in the first few coordination shells are included (Appendix and figures 4-5). Therefore, the properties that depend on the most collective modes may be studied independent of this choice. This is incontrast to the modes that affect the medium to high frequency motions. Therefore, in studies deriving information by relying on the superposition of a large number of modes, a cautious network construction is essential.

Network constructed by using Voronoi tessalations, on the other hand, fail to correctly define the local interactions while they successfully incorporate the long-range pairwise interactions. In particular, the mode distributions (figure 3a) and the spectral dimensions measured at the slow mode region (figure 3b) do not represent the experimentally and theoretically well-characterized shapes for proteins. Therefore, these network models will provide misleading information on the properties that rely mostly on local interactions (e.g. residue fluctuations, figure 4). On the other hand, they are expected to be very effective in forecasting properties that depend on a correct incorporation of the long-range contacts, as was recently demonstrated by their success in predicting the folding rates of two-state proteins (44).

## APPENDIX. Partitioning the Hessian into its essential and residual components.

We partition the Hessian into two parts:

$$H = H_* + H_r \qquad (A1)$$

We postulate that $H_*$ contains information due to the essential contacts of the nodes, whereas $H_r$ is the residual part where the interactions are added in a spherically symmetrical manner around the nodes beyond a certain $r_c$ value (figure 2). An alternative way of achieving the Hessian is through the use of an overall force balance around each node of the network as discussed in detail in reference (15). Therein, it is shown that the Hessian is obtained by the product of the $3N \times M$ direction cosine matrix, **B** with its transpose, for a system of $N$ nodes and a total of $M$ equivalent interactions between pairs of nodes;

$$H = BB^T \qquad (A2)$$

Inasmuch as the overall interactions may be written as the sum of the $B_*$ and $B_r$ matrices that contain the essential and the residual interactions, the Hessian may thus be expressed as

$$\begin{aligned} H &= (B_* + B_r)(B_* + B_r)^T \\ &= B_* B_*^T + B_* B_r^T + B_r B_*^T + B_r B_r^T \end{aligned} \qquad (A3)$$

We now denote the elements of the $B_*$ and $B_r$ matrices by cos $\alpha_{ij}$ and cos $\beta_{ij}$, respectively. Then, the elements of the terms in equation A3 may be calculated from,

$$(B_* B_*^T)_{ij} = \begin{cases} \sum_{k=1}^m \cos^2 \alpha_{ik} & i = j \\ \sum_{k=1}^m \cos \alpha_{ik} \cos \alpha_{jk} & i \neq j \end{cases} \qquad (A4)$$

$$(B_r B_r^T)_{ij} = \begin{cases} \sum_{k=1}^m \cos^2 \beta_{ik} \approx \pi & i = j \\ \sum_{k=1}^m \cos \beta_{ik} \cos \beta_{jk} \approx 0 & i \neq j \end{cases} \qquad (A5)$$

$$(B_r B_*^T)_{ij} = (B_* B_r^T)_{ji} = \begin{cases} \sum_{k=1}^m \cos \beta_{ik} \cos \alpha_{ik} = 0 & i = j \\ \sum_{k=1}^m \cos \beta_{ik} \cos \alpha_{jk} \approx 0 & i \neq j \end{cases} \qquad (A6)$$

The approximations in equations A5 and A6 follow from the assumption that the elements of the residual matrix, cos $\beta_{ij}$ are uncorrelated and span the entire range of values in [-$\pi$, $\pi$], and $m$ is large. These lead to the result,

$$H \approx B_* B_*^T + \pi I \qquad (A7)$$

Comparing equations A1 and A7 suggests that the modification to the essential part of the Hessian, due to a set of randomly added interactions, is through the addition of a constant. In the inverse operation, this leads to a modification of the eigenvalues in the range [0, $\pi$], whereas the eigenvectors are expected to remain unchanged (45).

In reality, this is an idealization of the residual matrix, and the eigenvalues are expected to display a distribution of values, peaking at $\pi$. To determine the extent to which the interactions above a threshold distance value approximate $H_r$, we decompose the Hessian of the protein set such that the interactions up to 17 Å are incorporated into the essential part, $H_*$, and those in the range 17 – 25 Å are considered in the residual, $H_r$. We verify that the eigenvalue distribution of the latter matrix peaks around $\pi$. Furthermore, using,

$$H = H_* + H_r = H_* \left( I + H_*^{-1} H_r \right) \qquad (A8)$$

we find that more than 50 % of the eigenvalues of $H_*^{-1} H_r$ is below 1 so that the main features of the essential part of the Hessian, in particular the slow modes, remain unchanged in the total Hessian.

**ACKNOWLEDGEMENTS.** We thank Tural Aksel for bringing up the observation that large cut-off distances may improve the predictions on the residue fluctuations. İİ thanks Deniz Turgut for his help in parts of the computations.

**FIGURE CAPTIONS**

**Figure 1.** (a) The negative of the resultant vectors acting on the nodes, $-\mathbf{Q}_i$, exemplified by a 54 residue protein (PDB code: 1enh). The length of each red vector is proportional to the cut-off distance used in network construction, $r_c$, the shortest at 7 Å and the longest at 15 Å. The yellow vector is the resultant obtained from the networks obtained from the Voronoi tessalations. (b) Part of the helix marked by the square in (a) is magnified; "exterior" refers to the solvent contacting part of the helix, and "interior" marks the side facing the core of the protein.

**Figure 2.** (a) Radial and angular distribution functions (left y-axis: RDF; right y-axis: ADF) obtained by averaging over 595 proteins. (b) ADFs computed separately for the core and surface residues for a subset of 60 proteins.

**Figure 3.** (a) The change of the density of vibrational modes, $g(\omega)$, with the cut-off distance, $r_c$, used in network construction. The main figure displays the results for $r_c$ in the first ($r_c = 7$ Å) to above the fourth coordination shell range (up to 16 Å). Also shown, in dashed lines, is the frequency distribution of the Voronoi tessalated networks. The inset displays the results for very large $r_c$ values (up to 30 Å). The data is an average over a set of 26 proteins in the size range of $150 \pm 10$ residues. (b) Spectral dimension, $d_s$, of the networks, obtained from power law best-fits to the cumulative density of modes, $G(\omega) \propto \omega^{d_s}$ for the first 70 modes in each set of data. Goodness of fit is 0.98 or better in all cases. The thin dashed lines are included to guide the eye for the cross-over in the rate of change of $d_s$ with $r_c$. Also indicated on the figure are the $d_s$ of the Voronoi tessalated networks that occur at ca. 1.0, and the theoretical limit at $d_s = 3$ when all nodes are interconnected ($r_c \rightarrow \infty$).

**Figure 4.** (a) Comparison of the X-ray B-factors (gray, middle curve) with fluctuation profiles predicted from various models (at $r_c = 8$, 15, 25 Å and network construction with Voronoi tessalations) for the 268 residue achromobacter lyticus protease (PDB code: 1arb). (b) Pearson correlation coefficients at a wide range of cut-off distances for the same protein; those that correspond to the detailed fluctuation profiles of figure 4a are shown with filled circles and that with the cut-off free Voronoi tessalated model is marked by the dashed line.

**Figure 5.** The displacement profiles of adenylate kinase in unbound and bound forms (PDB codes: 4ake and 1ake, respectively). The experimental displacements are shown in gray. Predictions from the relative magnitudes of the eigenvector corresponding to the slowest mode obtained at $r_c = 8$, 10, 15 Å are shown in black. The latter curves are displaced to guide the eye, and their zero baselines are marked by the dotted curves. The Pearson correlation between the experimental curve and each of the predictions is 0.9.

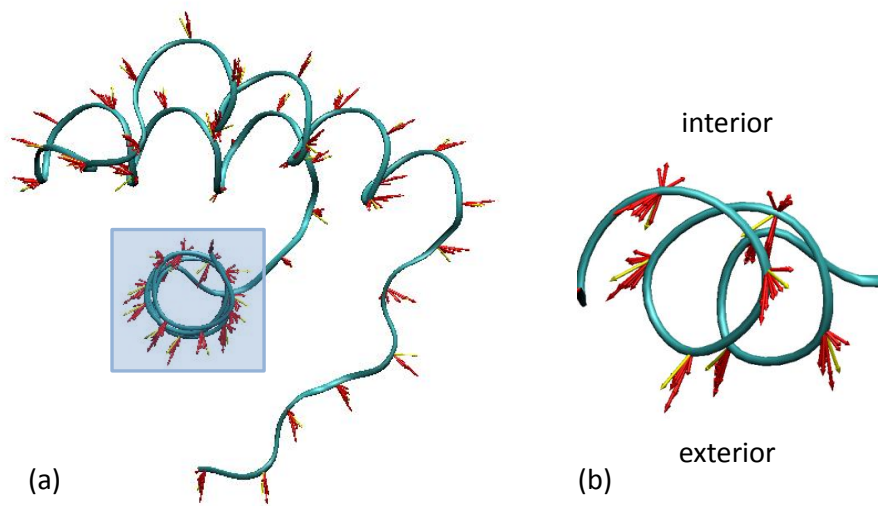

(a)

(b)

interior

exterior

*Figure 1*

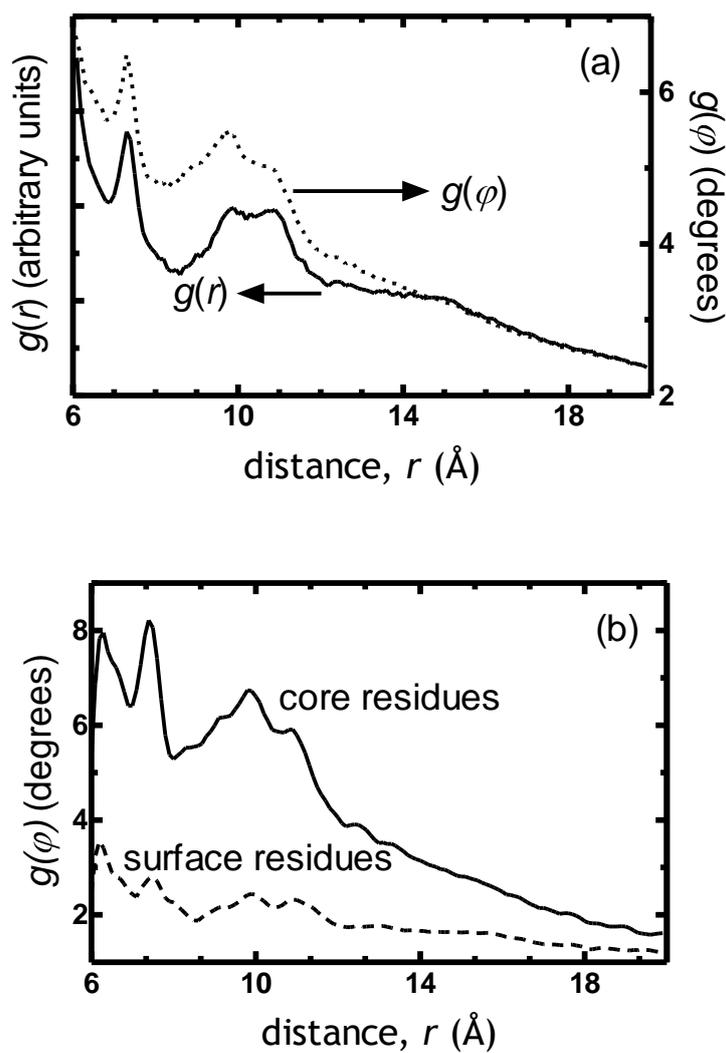

(a)

(b)

*Figure 2*

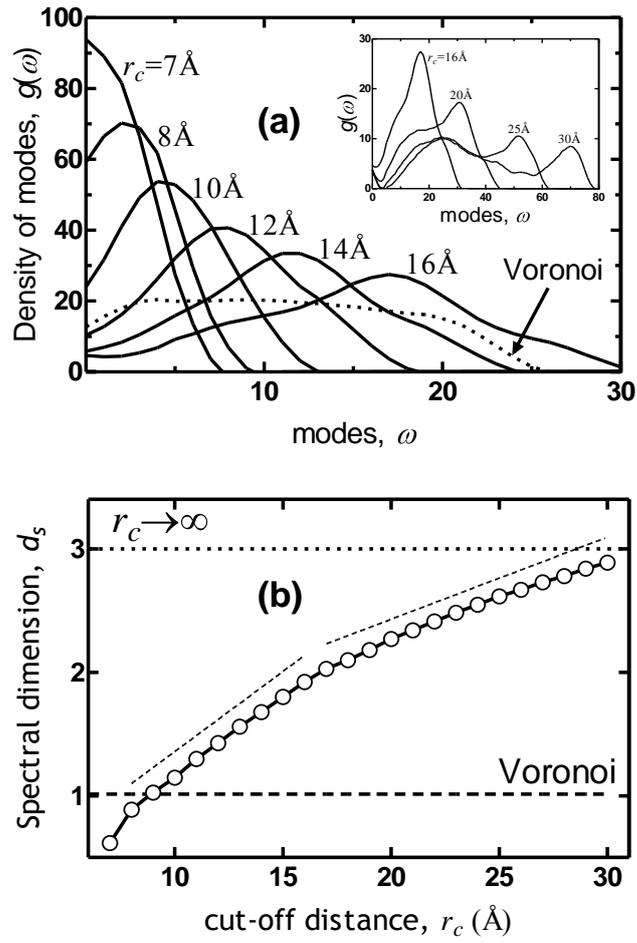

*Figure 3*

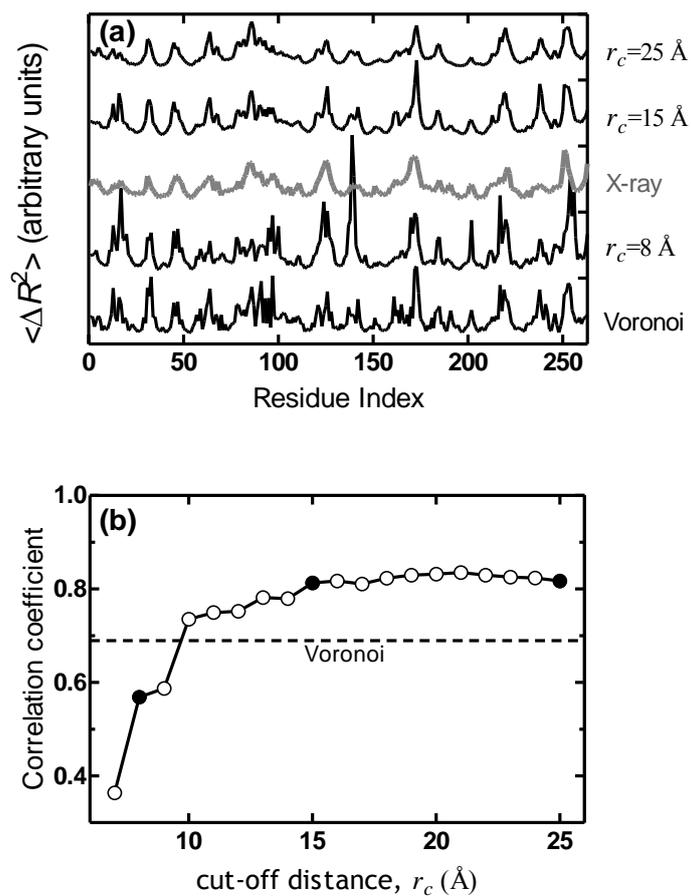

*Figure 4*

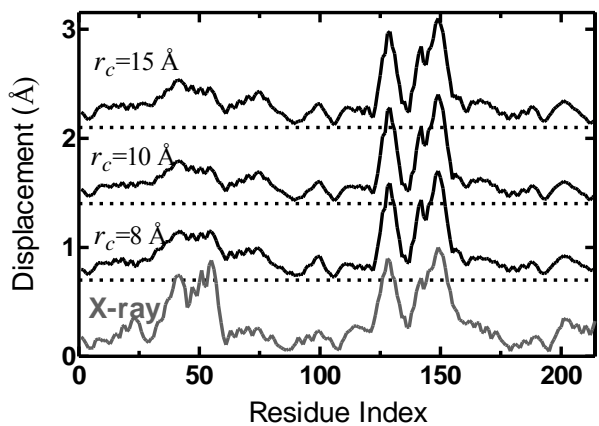

*Figure 5*